\documentclass[12pt,reqno]{amsart}

\usepackage[T1]{fontenc}
\usepackage{graphicx}
\usepackage{cite,enumerate}
\usepackage{float}
\usepackage{amsthm}
\usepackage{amssymb}
\usepackage{color}
\definecolor{MyLinkColor}{rgb}{0,0,0.4}

\newcommand{\e}{\varepsilon}

\newcommand{\Beq}{\begin{equation}}
\newcommand{\Eeq}{\end{equation}}
\newcommand{\BS}{\begin{subequations}}
\newcommand{\ES}{\end{subequations}}
\newcommand{\Beqn}{\begin{equation*}}
\newcommand{\Eeqn}{\end{equation*}}
\newcommand{\Beqa}{\begin{eqnarray}}
\newcommand{\Eeqa}{\end{eqnarray}}
\newcommand{\Beqan}{\begin{eqnarray*}}
\newcommand{\Eeqan}{\end{eqnarray*}}

\newcommand{\disp}{\displaystyle}

\newtheorem{thm}{Theorem}[section]
\newtheorem{prop}[thm]{Proposition}

\theoremstyle{remark}

\numberwithin{equation}{section}

\title[]{Instability of equatorial water waves with an underlying current}

\author[F. Genoud]{Fran\c{c}ois Genoud}
\address{Faculty of Mathematics, University of Vienna,
Vienna, Austria}
\email{francois.genoud@univie.ac.at}
\author[D. Henry]{David Henry}
\address{School of Mathematical Sciences, University College Cork, Cork, Ireland.}
\email{d.henry@ucc.ie}

\subjclass[2010]{Primary: 76B15; Secondary: 76B47, 35B50, 26E05}
\keywords{Geophysical flows, instability, currents.}

\usepackage[colorlinks=true,linkcolor=MyLinkColor,citecolor=MyLinkColor]{hyperref}

\begin{document}

\begin{abstract}
In this paper we use the short-wavelength instability  approach to derive an instability threshold for exact trapped equatorial waves propagating eastwards in the presence of an underlying current.
\end{abstract}

\maketitle

\section{Introduction}
In this paper we consider the hydrodynamical stability of a three-dimensional exact solution of the governing equations for  equatorial geophysical  water waves  which was recently presented in \cite{Hen2013}. This solution is explicit in the Lagrangian formulation, and it prescribes trapped equatorial waves which propagate eastwards  in the presence of a constant underlying current. Physically, the equator acts  as a natural waveguide, leading to trapped zonal waves which decay exponentially away from the equator, cf. the discussions in \cite{ConJGR,Fed}.  Large-scale currents and wave-current interactions play a major role in the geophysical dynamics of the equatorial region, and it has been proposed that the interplay between equatorial currents in the ocean and atmosphere is one of the major generating mechanisms for El Ni\~no and La Ni\~na phenomena, cf. \cite{Iz} and the discussions in \cite{ConJGL,ConDIE,HenMat}. 
 In \cite{Mol}, Mollo-Christensen introduced a current-like term into Gerstner's solution for gravity waves in order to describe billows between two fluids, and this idea was extended to the geophysical  setting in \cite{Hen2013}.

Hydrodynamical stability is an important area of fluid mechanics which has been addressed systematically since the nineteenth century, following the seminal work of Kelvin, Helmholtz, Rayleigh and Reynolds, cf. \cite{Dr,DR,Fried2} for an overview of the mathematical approach to hydrodynamic stability. From a mathematical viewpoint, establishing the hydrodynamical stability or instability of a flow is  difficult, given that the fully nonlinear governing equations for fluid motion are highly intractable, and accordingly there exist only a handful of explicit exact solutions. Among these is the famous Gerstner's wave \cite{ConGer1,Ger,HenGer}, an exact solution that is explicit in the Lagrangian formulation. Recently, Gerstner-type solutions have been derived and adapted to model a number of different physical and geophysical scenarios  \cite{ConGer2,ConJGR,R2,ConJPO,Mat,St} with explicit exact solutions in a Lagrangian framework \cite{Ben}.

The  notion of hydrodynamical stability runs as follows. We suppose that a given fluid motion is perturbed by an infinitesimal disturbance, and the question of interest is as to how that infinitesimal perturbation evolves as time progresses. Physically, the question of hydrodynamic stability is important for numerous reasons. For instance, unstable flows cannot  be observed in practice since they are rapidly destroyed by any minor perturbations or disturbances. 
In this paper we employ the short-wavelength instability method to prove that if the wave steepness exceeds a certain value, then the equatorial water waves presented in  \cite{Hen2013} are unstable under short wavelength perturbations.  The short-wavelength instability method, which was independently developed by the authors of \cite{Bay,Fried1,LH},  examines how a localised and rapidly-varying infinitesimal perturbation will evolve by way of a system of ODEs. For certain solutions which have an explicit Lagrangian formulation, it transpires that the short wavelength instability analysis is remarkably elegant, and the criteria for instability \eqref{exp}--\eqref{exppos} takes on a tangible and explicit formulation in terms of the wave steepness. This was first established in \cite{Leb} for Gerstner's solution to the gravity water wave problem, and recently for geophysical flows in \cite{ConstGer} and for edgewaves in \cite{R1}. The wave-steepness instability criteria which applies to the equatorially trapped wave with an underlying constant current is presented in Proposition~\ref{prop} below.

\section{Governing equations}
In a reference frame with the origin located at a point on earth's surface and rotating with the earth, we take the $x-$axis to be the longitudinal direction  (horizontally due east), the $y-$axis to be the latitudinal direction (horizontally due north) and the $z-$axis to be vertically upwards. We take the earth to be a perfect sphere of radius $R=6378km$, which has a constant rotational speed of $\Omega=73\cdot10^{-6}rad/s$, and $g=9.8ms^{-2}$ is the standard gravitational acceleration at the earth's surface. The governing equations for geophysical ocean waves \cite{CR,Gal} are given by 
\begin{subequations}\label{full}
\begin{align}
u_t+ uu_x+vu_y+wu_z +2 \Omega w\cos \phi-2\Omega v \sin \phi &=-\frac 1 \rho P_x,   \\  
v_t+uv_x+vv_y+wv_z+2\Omega u \sin \phi&=- \frac 1 \rho P_y,  \\
w_t+uw_x+vw_y+ww_z-2\Omega u \cos \phi&=- \frac 1 \rho P_z-g,
\end{align}
\end{subequations}
together with the equation of incompressibility
\begin{subequations}\label{Gov}
\Beq\label{in}
\nabla \cdot{\bf U} =0.
\Eeq
Here  ${\bf U}=(u,v,w)$ is the velocity field of the fluid, the variable $\phi$ represents the latitude,  $\rho$ is the density of the fluid (which we take to be constant), and $P$ is the pressure of the fluid. For latitudes within $5^{\circ}$ of the equator we operate within the framework of the $\beta-$plane approximation \cite{CR} of the governing equations \eqref{full} 
\begin{subequations}\label{beta}
\begin{align}
u_t+ uu_x+vu_y+wu_z +2 \Omega w-\beta y v &=-\frac 1 \rho P_x,   \\  
v_t+uv_x+vv_y+wv_z+\beta y u&=- \frac 1 \rho P_y,  \\
w_t+uw_x+vw_y+ww_z-2\Omega u &=- \frac 1 \rho P_z-g,
\end{align}
\end{subequations}
where  $\beta=2\Omega/R=2.28\cdot 10^{-11}m^{-1}s^{-1}$. 
The boundary conditions for the fluid on the free-surface $\eta$ are given by
\begin{align}
\label{k} w&=\eta_t+u\eta_x +v\eta_y,   \\
\label{p}  P&=P_{0}  \qquad \mathrm{on}\ y=\eta(x,y,t), 
\end{align}
where $P_{0}$ is the constant atmospheric pressure. The kinematic boundary condition on the surface simply states that all surface particles remain confined to the surface. For  trapped equatorial waves the wave surface profile decays in the latitudinal direction away from the equator, and we assume the water to be infinitely deep. Since we model waves which propagate over a flow with a constant underlying current, the flow converges rapidly  with increasing depth to a constant uniform current 
\Beq \label{lim}
(u,v)\rightarrow(-c_0,0) \  \mbox{ as } \ y\rightarrow -\infty.  
\Eeq
\end{subequations}
The direction of the current has a bearing on the dynamics of the exact solution that we outline below, and additionally it plays a role in either increasing or decreasing the likelihood of instability of the flow with respect to short wavelength perturbations.

\section{Exact solution of \eqref{beta}}
Recently, in \cite{Hen2013}, an exact solution to the $\beta-$plane equations \eqref{beta} with a constant underlying current was presented. The solution represents steady waves travelling in the longitudinal direction due east, which have a constant  speed of propagation  $c>0$, in the presence of a constant underlying current of strength $c_0$.  The Eulerian coordinates of fluid particles $(x,y,z)$ are expressed as functions of the time $t$ and the Lagrangian labelling variables $(q,r,s)$, for $q\in \mathbb R$, $r\in(-\infty,r_0]$ where  $r_0<0$, and $s$ in an interval which depends on the direction and size of $c_0$, as follows:
\begin{subequations}\label{lvara}
\begin{align}
x&=q-c_0 t-\frac 1k e^{k[r-f(s)]}\sin{[k(q-c t)]},
\\
y&=s,\\
z&=r+\frac 1k e^{k[r-f(s)]}\cos{[k(q-c t)]}.
\end{align}
\end{subequations}
Here $k$ is the wavenumber and the function $f(s)$  takes the form
\Beqn\label{feq}
f(s)=\frac{c \beta}{2\gamma}s^2,
\Eeqn
 where 
\Beqn \label{GM}\gamma=2\Omega c_0 +g>0\Eeqn
for all physically plausible values of the current $c_0$. The function $f(s)$ ensures a latitudinal decay in the fluid motion  away from the equator when $c>0$. 
Defining
\[
\chi=k\left(r-f(s)\right),\ \theta=k(q-c t),\] 
the determinant of the Jacobian of the transformation \eqref{lvara} is $1-e^{2\chi}$, which is time independent. Thus it follows that the flow defined by \eqref{lvara} must be volume preserving, ensuring that \eqref{in} holds in the Eulerian setting \cite{Ben}. In order for the transformation \eqref{lvara} to be well-defined, and  to  ensure that our flow has the appropriate decay properties (in both the vertical and the latitudinal directions), it is necessary that 
\Beqn \label{ineq}
r-f(s)\leq r_0<0.
\Eeqn We note that this relation implies that $c>0$ for our flow, and so the waves must propagate in an easterly direction. As we are modelling trapped equatorial waves, we take $v\equiv 0$ throughout the fluid, and we can express \eqref{beta} as
\begin{subequations}\label{Gov1}
\begin{align}\label{Eq1}
\frac{Du}{Dt}+2\Omega w   &= -\frac{1}{\rho}P_x, \\
\frac{Dv}{Dt}+\beta y u   &= -\frac{1}{\rho}P_y, \\
\frac{Dw}{Dt}-2\Omega u   &= -\frac{1}{\rho}P_z-g,
\end{align}
\end{subequations} 
where $D/Dt$ is the material derivative. The 
velocity gradient tensor
\begin{align}\label{transp}
\nabla{\bf U}
=\left(
\begin{array}{ccc}
\frac{\partial u}{\partial x} & \frac{\partial v}{\partial x}  & \frac{\partial w}{\partial x}  \\
\frac{\partial u}{\partial y} & \frac{\partial v}{\partial y}  & \frac{\partial w}{\partial y}  \\
\frac{\partial u}{\partial z} & \frac{\partial v}{\partial z}  & \frac{\partial w}{\partial z}  
\end{array}
\right)
=\frac{c k e^{\chi}}{1-e^{2\chi}}\left(
\begin{array}{ccc}
{-\sin{\theta}} & 0 & \cos{\theta}+e^{\chi} 
\\
f_s(e^{\chi}-\cos \theta) & 0 & -f_s\sin \theta  
\\
-e^{\chi}+\cos \theta & 0 & \sin \theta 
 \end{array}
\right),
\end{align}
and so the vorticity  is given by $\omega=(w_y-v_z,u_z-w_x,v_x-u_y)$
\Beqan
= \left(
-s\frac{kc^2\beta}{g}\frac{e^{\chi}\sin \theta}{1-e^{2\chi}}, -\frac{2kc e^{2\chi}}{1-e^{2\chi}},
s\frac{kc^2\beta}{g}
\frac{e^{\chi}\cos \theta-e^{2\chi}}{1-e^{2\chi}}
\right).
\Eeqan
The motion prescribed by \eqref{lvara} satisfies \eqref{Gov1} when the pressure function is 
\Beqn \label{Pa}
P=\rho \gamma \left(\frac{e^{2\chi}}{2k}-r+\frac {c_0} c f(s)\right)+P_0 -\rho g \left(\frac{e^{2kr_0}}{2k}-r_0\right),
\Eeqn 
and when the dispersion relation holds:
\Beq \label{aa}
c= \frac{\sqrt{\Omega^2+k(2\Omega c_0 +g)}-\Omega}{k}, \quad  c_0\neq c.
\Eeq 
The flow determined by \eqref{lvara} satisfies the governing equations \eqref{beta}, and the free-surface $z=\eta(x-ct,y)$ is defined parametrically at fixed latitudes $y=s$ by setting $r=r(s)\leq r_0$,
where $r(s)$ is the unique solution of the  equation
\Beqn \label{sol}
\frac{e^{2k[r-\frac{c\beta}{2\gamma}s^2]}}{2k}-r+ \frac{c_0\beta}{2\gamma}s^2-\frac{e^{2kr_0}}{2k}+r_0=0.
\Eeqn
When $c_0\leq 0$, which corresponds to a following current, the above equation has a unique solution for all $s\in \mathbb R$, and the solution \eqref{lvara} defines a travelling trapped equatorial wave. For $c_0>0$, an adverse current, the solution exists only for $s$ in a restricted region, cf. \cite{Hen2013}.  The surface wave prescribed by \eqref{lvara}, for fixed values of $s$ and $t$, is an upside down trochoid, cf. \cite{ConBook,ConJGR,HenGer}. Also, the steepness of the wave-profile, defined to be half the amplitude multiplied by the wavenumber, is given by
\[
\tau(s)=e^{\chi},
\] which is  maximum $\tau_0=e^{kr_0}$ at the equator. In the absence of a current, or in a reference frame moving with uniform speed $c_0$, we can see from \eqref{lvara} that the  particle trajectories for the underlying flow are closed circles in a fixed latitudinal plane. The effect of the current in \eqref{lvara} is to transport the particles in the zonal direction with uniform speed $c_0$. The existence of closed particle paths is typical of Gerstner-type waves, and it is a phenomenon which does not apply to most irrotational water waves. In the setting of both finite \cite{R2,R3} and infinite \cite{R4} depth Stokes waves, and for linear irrotational waves \cite{p4}, the particle trajectories are in fact not closed.  

\section{Instability analysis}
The main result of this paper may be stated as follows:
\begin{prop}\label{prop}
The equatorial waves propagating eastward over a constant underlying current, as prescribed by \eqref{lvara}, are unstable to short wavelength perturbations if the steepness of the wave
\begin{equation}\label{instcond2}
e^{kr_0}>\frac{3\Omega+\sqrt{\Omega^2+k(2\Omega c_0+g)}}
{\Omega+3\sqrt{\Omega^2+k(2\Omega c_0+g)}} \gtrapprox \frac 13.
\end{equation}
\end{prop}
The proof of Proposition \ref{prop} will be presented in the remainder of this section. We  employ the short-wavelength instability method, which examines the evolution of a localised and rapidly-varying infinitesimal perturbation represented at time $t$ by the wave packet  
\begin{equation}\label{wave}
{\bf u}({\bf X},t)=\e {\bf b}({\bf X},{\bf \xi_0},{\bf b_0},t)e^{i\Phi({\bf X},{\bf \xi_0},{\bf b_0},t)/\delta}.
\end{equation}
Here ${\bf X}=(x,y,z)$, $\Phi$ is a scalar function, and at $t=0$ we have
\[
\Phi({\bf X},{\bf \xi_0},{\bf b_0},0)={\bf X }\cdot {\bf \xi_0}, \quad {\bf b} ({\bf X},{\bf \xi_0},{\bf b_0},0)
={\bf b_0}({\bf X},{\bf \xi_0}).
\] 
The normalised wave vector ${\bf \xi_0}$  is subject to the transversality condition ${\bf \xi_0}\cdot{\bf b_0}=0$, 
and ${\bf b_0}$ is the normalised amplitude of the short-wavelength perturbation of the flow which has 
the velocity field 
${\bf U}({\bf X})\equiv (u \ v \ w)^T(x,y,z)$.  Then the evolution in time of $\bf X$, of the perturbation amplitude $\bf b$, and of the wave vector ${\bf \xi}=\nabla \Phi$, is governed at the leading order in 
the small parameters $\epsilon$ and $\delta$ by the system of ODEs
\Beq\label{pertsyst}
\begin{cases}
\dot{\bf X}={\bf U }({\bf X},t), \\
\dot{{\bf \xi}}=-(\nabla{\bf U})^T{\bf \xi}, \\
\dot{{\bf b}}=-L{\bf b}-(\nabla{\bf U}){\bf b}+
([L{\bf b}+2(\nabla{\bf U}){\bf b}]\cdot{\bf \xi})\disp\frac{{\bf \xi}}{|{\bf \xi}|^2},
\end{cases}
\Eeq
with initial conditions 
\Beqn
\label{IC}
{\bf X}(0)={\bf X_0},\ {\bf \xi}(0)={\bf \xi_0}, \ {\bf b}(0)={\bf b_0}.
\Eeqn Here $(\nabla{\bf U})^T$ is the transpose of the velocity gradient tensor \eqref{transp} and, for the system defined by \eqref{lvara},  $L=L({\bf X})$ is given by 
\[
L=
\begin{pmatrix}
0 & -\beta y & 2\Omega \\ \beta y & 0 & 0 \\ -2\Omega & 0 & 0
\end{pmatrix},
\] cf. \cite{ConstGer} for details.

The system of ODEs \eqref{pertsyst} describing the evolution of a rapidly-varying perturbation was independently derived  in \cite{Bay,Fried1,LH}.
The instability criterion for Lagrangian flows  for which ${\bf X}(0)=\bf X_0$ is determined by the exponent
\begin{equation} \label{exp}
\Lambda({\bf X_0})=\limsup_{t \rightarrow \infty}\frac 1 t 
\ln \left( \sup_{|{\bf \xi_0}|=|{\bf b_0}|=1,{\bf \xi_0}\cdot {\bf b_0}=0}
\left\{ |{\bf b}({\bf X},{\bf \xi_0},{\bf b_0},t)| \right\} \right).
\end{equation}
If 
\begin{equation} \label{exppos}
\Lambda({\bf X_0})>0 
\end{equation}
for a given fluid trajectory then particles become separated at an exponential rate, and accordingly the flow is unstable \cite{Fried}. Therefore, establishing \eqref{exppos} provides us with a criterion  for instability of a flow.
For certain solutions \cite{ConstGer,Leb} which have an explicit Lagrangian formulation, it transpires that the short wavelength instability analysis is remarkably elegant, and the criterion for instability \eqref{exppos} takes on a tangible and explicit formulation in terms of the wave steepness. We now derive this instability criterion for the flow determined by \eqref{lvara}. 

We choose the latitudinal wave vector  ${\bf \xi_0}=(0 \ 1 \ 0)^T$, and from \eqref{transp} we then have ${\bf \xi}(t)=(0 \ 1 \ 0)^T$
for all $t\geq0$. It follows that the evolution of ${\bf b}=(b_1,b_2,b_3)$ is governed by
\begin{equation}\label{3comp}
\begin{cases}
\dot{b_1}=\disp\beta s b_2 - 2\Omega b_3 + \frac{kce^\chi\sin\theta}{1-e^{2\chi}}b_1
		-\frac{kc^2\beta s e^\chi(e^\chi-\cos\theta)}{(2\Omega c_0+g)(1-e^{2\chi})}b_2
		+\frac{kce^\chi(e^\chi-\cos\theta)}{1-e^{2\chi}}b_3,\\
\dot{b_2}=0, \\
\dot{b_3}=\disp2\Omega b_1-\frac{kce^\chi(e^\chi+\cos\theta)}{1-e^{2\chi}}b_1
		+\frac{kc^2\beta s e^\chi\sin\theta}{(2\Omega c_0+g)(1-e^{2\chi})}b_2
		-\frac{kce^\chi\sin\theta}{1-e^{2\chi}}b_3.
\end{cases}
\end{equation}
Noting that the choice $b_2(0)=0$ implies $b_2(t)=0$ for all $t\geq0$, and accordingly ${\bf \xi}(t)\cdot {\bf b}(t)=0$, the system \eqref{3comp} reduces to
\Beq\label{2comp}
\dot B=
\left(\begin{array}{cc}
\frac{kce^\chi\sin\theta}{1-e^{2\chi}}	& -2\Omega+\frac{kce^\chi(e^\chi-\cos\theta)}{1-e^{2\chi}} \\
2\Omega -\frac{kce^\chi(e^\chi+\cos\theta)}{1-e^{2\chi}} & 		-\frac{kce^\chi\sin\theta}{1-e^{2\chi}}
\end{array}
\right)B, 
\Eeq where $B=\left(\begin{array}{c} b_1 \\ b_3 \end{array}  \right)$. This system is nonautonomous, however the change of variables induced by the matrix
\[
P=\left(\begin{array}{cc}
\cos\left(kct/2 \right)	& \sin\left(kct/2 \right) \\
-\sin\left(kct/2 \right) & \cos\left(kct/2 \right)
\end{array}
\right)
\] transforms the planar system \eqref{2comp} to an autonomous system for $Q=P^{-1}B$, 
\[
\frac{d}{dt}Q(t)=DQ(t),
\] where 
\[
D=
\begin{pmatrix}
\frac{kce^{\chi}}{1-e^{2\chi}}\sin(kq) & -\frac{kce^{{\bf \chi}}}{1-e^{2\chi}}\cos(kq)-2\Omega+\frac{kce^{\chi}}{1-e^{2\chi}}-\frac{kc}2
\\ -\frac{kce^{\chi}}{1-e^{2\chi}}\cos(kq)+2\Omega-\frac{kce^{\chi}}{1-e^{2\chi}}+\frac{kc}2 & -\frac{kce^{\chi}}{1-e^{2\chi}}\sin(kq) 
\end{pmatrix}.
\]
Since $B=PQ$, and $P$ is periodic with time, we deduce that the short-wavelength rapidly-varying perturbation ${\bf u}$, defined in \eqref{wave}, grows exponentially with time if $D$ has a positive eigenvalue, which occurs if 
\begin{equation*}\label{instcond}
e^\chi>\frac{4\Omega+kc}{4\Omega+3kc}.
\end{equation*}
Using the dispersion relation \eqref{aa}, this gives us condition \eqref{instcond2}. 
Therefore, if the steepness  of the wave is sufficiently large at the equator, the localised small perturbation \eqref{wave} of  grows at an exponential rate and the flow is consequently unstable. 
We note that since $\Omega\ll 1$ the left-hand side of \eqref{instcond2} is of the form 
\begin{equation*}
\frac{3\tilde \epsilon +1}
{\tilde \epsilon+3} \gtrapprox \frac 13,
\end{equation*}
where 
\[
\tilde \epsilon=\frac{\Omega}{\sqrt{\Omega^2+k(2\Omega c_0+g)}}\ll 1.
\] In particular, we deduce that an adverse current with $c_0>0$ favours instability
in the sense that the threshold on the steepness for the wave to be unstable is decreased
compared to the case without current. Conversely, this threshold is increased
by a following current with $c_0<0$.

\subsection*{Acknowledgment}
The support of the ERC Advanced Grant ``Nonlinear studies of water
flows with vorticity'' is acknowledged.

\end{document}